# Ungapped magnetic excitations beyond Hidden Order in URu$_{2-x}$Re$_x$Si$_2$

Nicholas P. Butch[a,b]*, J. A. Rodriguez-Rivera[a,c], and M. Brian Maple[d]

[a]*NIST Center for Neutron Research, Gaithersburg, MD 20899 USA;* [b]*Center for Nanophysics and Advanced Materials, University of Maryland, College Park, MD 20742 USA;* [c]*Department of Materials Sciences, University of Maryland, College Park, Maryland 20742, USA* [d]*Department of Physics, University of California, San Diego, La Jolla, CA 92093 USA*

*corresponding author: nbutch@umd.edu

# Ungapped magnetic excitations beyond Hidden Order in $URu_{2-x}Re_xSi_2$

We use inelastic neutron scattering measurements to show that the energy gap in the magnetic excitations of $URu_2Si_2$, induced by the Hidden Order transition, is closed by Re substitution. The magnetic excitations remain ungapped in compositions where the specific heat anomaly associated with Hidden Order is no longer observed, which means that the entropy associated with Hidden Order is tied mostly to the magnetic gapping. Further, the onset of ferromagnetic order does not gap the excitations, reflecting the fact that Re substitution does not dramatically affect the Kondo lattice hybridization.

**Introduction**

The Hidden Order (HO) transition of the heavy fermion compound $URu_2Si_2$ is characterized by a large jump in the specific heat at $T_{HO}$ = 17.5 K [1-3]. To date, there is no conclusive determination of the microscopic order parameter that is responsible for this phase transition [4]. However, there is no evidence that HO is dipolar or quadrupolar, leading to current theoretical thinking that higher-order multipolar states are likely candidates [5,6].

The energy gap structure of the magnetic excitations has a complicated wavevector Q-dependence in the HO phase [7,8], which is typically summarized by quoting the energy gap values E at minima $E_Z$ = 2 meV at $Q_Z$ = (1,0,0) and symmetrically equivalent Q, and $E_\Sigma$ = 5 meV at $Q_\Sigma$ = (0.6,0,0) and symmetrically equivalent Q [9]. The Q-dependence of the spin excitations is consistent with interband scattering in a correlated, hybridized band structure [10]. Therefore, the gapping of the spin excitations reflects a modification of the band structure at $T_{HO}$.

A way to study the relationship between the iconic specific heat anomaly and the gap in the spin excitations is to systematically suppress the HO transition. This can be done via Re substitution in the series $URu_{2-x}Re_xSi_2$ [11], which does not induce competing antiferromagnetic AFM order at low concentrations [12], as seen in the case of Rh, and yields a well-separated ferromagnetic FM phase at sufficiently high Re concentration [13]. Because the specific heat anomaly can only be tracked to x = 0.10 [14], the phase transition is either discontinuous or the HO phase becomes gapless [15].

By studying inelastic neutron scattering on URu$_{2-x}$Re$_x$Si$_2$ samples with Re concentrations near the limit of HO stability, we demonstrate that although signatures of Kondo lattice physics remain, when HO is suppressed, the magnetic excitation gap associated with HO does not open. This suggests that the majority of the entropy associated with the HO transition is due to Fermi surface reconstruction. Accordingly, in the FM state where there is no large specific heat anomaly, no Fermi surface reconstruction occurs.

**Methods**

Samples were cut from single crystals of URu$_{2-x}$Re$_x$Si$_2$ that were synthesized via the Czochralski technique in a multi-arc furnace from arc-melted polycrystals. These single crystals were previously characterized magnetically and via specific heat and transport measurements [14-16]. Inelastic neutron scattering measurements were performed on samples of x = 0.12, 0.15, and 0.25 in the (h k 0) plane using the MACS cold neutron spectrometer [17] and a helium-flow cryostat.

**Results**

The magnetic excitation spectrum of URu$_2$Si$_2$ in the HO phase is dominated by gapped basal-plane excitations with an approximate bandwidth of 10 meV [7]. Prominent branches are centered at commensurate $Q_Z$ and incommensurate $Q_\Sigma$. The Z-point excitations are narrow in Q, with a minimum energy of 2 meV, whereas the incommensurate excitations are rather extended, following the Brillouin zone edge, and have a minimum energy of 5 meV. At higher temperatures, outside of the HO phase, the energy gaps close, although the Q structure is maintained [10]. This overall behaviour has been attributed to interband excitations from hybridized electronic bands, as opposed to magnon scattering.

Starting at the edge of the HO state, we compare the magnetic excitations to the parent. The $x = 0.12$ sample shows no specific heat anomaly or transport signature associated with a transition into HO, which may be a sign that the HO phase is gapless here [15]. The Q-dependence of the magnetic excitations detected by inelastic neutron scattering is mapped in Fig. 1a. Excitations are centered at $Q_Z$ and $Q_\Sigma$, as in $URu_2Si_2$. The crucial difference is that down to 1.4 K, there is no indication of the opening of an energy gap (Fig. 1b-d). There is, however, a notable temperature dependence, as the scattered intensity increases on cooling from 10 K to 1.4 K (Fig. 1b). This increase is opposite the trend expected naively for the thermal population of magnetic excitations, implying that the spin fluctuations strengthen at low temperatures. As in the paramagnetic state in $URu_2Si_2$, the overall Q-energy dependence of $x = 0.12$ is not strongly temperature-dependent at low temperature(Fig. 1c,d). Relative to $URu_2Si_2$, the commensurate excitations appear weaker relative to the incommensurate excitations [10].

We next explore the effects of higher Re concentration. Based on magnetization scaling, a FM quantum critical point exists at $x = 0.15$ [16]. Although it is not yet firmly established whether $x = 0.12$ supports a gapless HO state [15], $x = 0.15$ is outside of the HO state, and exhibits no magnetic order. Yet, there is no discernible difference between the magnetic excitations at 1.4 K in $x = 0.15$ (Fig. 2a,b) and those in $x = 0.12$ (Fig. 1). This fact is consistent with the nearly identical low-temperature specific heat measured for both values of x [15].

Well inside the FM phase in x = 0.25, the magnetic excitations remain remarkably similar to those observed at lower x, although the relative intensity at Qz weakens somewhat (Fig. 2 c,d). The inelastic neutron scattering data are collected at 1.4 K in the FM state, well below the Curie temperature of 3.1 K. The similarity of the excitation spectrum to that in the paramagnetic state shows that the FM order does not open an energy gap. This observation is consistent with the lack of discernible specific heat anomalies at the Curie temperature and the overall similarity of the low-temperature specific heat of these small-moment, itinerant ferromagnets [15].

**Discussion**

In $URu_2Si_2$, the incommensurate magnetic excitations at $Q_\Sigma$ are due to interband scattering that is a signature of a heavily renormalized band structure [10]. The continuing presence of these excitations for all measured Re concentrations indicates that these materials are all heavy fermion metals containing similar renormalized band structures that arise from hybridization between f-states and light conduction bands for temperatures lower than 70 K or so. This means that the stability of different electronic ground states: FM, AFM, and HO, is associated with very minor changes to the electronic structure, as reflected in the rather similar transport, magnetic, and thermodynamic properties of these materials. Changes in possible Fermi surface nesting, or other wavevector-sensitive instabilities, are therefore unlikely to be directly responsible for the tuning of the different ordered phases by chemical substituion or moderate applied pressure [18].

Another consequence of the interband excitations pertains to earlier determination of quantum critical scaling [19]. The present measurements on single crystal samples clarify that the magnetic excitations previously observed in polycrystalline samples are not due to AFM fluctuations, but rather to the same interband transitions identified here, with peak intensity at $Q_\Sigma$. In fact, the temperature dependence reported for a polycrystalline sample of x = 0.20 matches that at $Q_\Sigma$ in the parent material [7].

Our observations also allow us some further insight into the HO symmetry. There is clearly a correlation between lattice symmetry breaking and the presence of a magnetic energy gap in the ground state. The x = 0.25 sample exhibits no energy gap in the FM state, where there is no lattice symmetry breaking, whereas AFM order opens a large gap in the presence of broken lattice symmetry [20]. Because the HO phase also opens an energy gap, it is tempting to surmise that the HO parameter breaks lattice symmetry like the AFM phase, perhaps with the same ordering vector Qz, as proposed for theories having coupled HO/AFM order parameters. The absence of the Qz excitations in the x = 0.25 data (Fig. 2 c,d) then leads to the conclusion that FM order becomes stable when competing AFM correlations are tuned away. We note that this simple picture is not yet complete. For example, it remains to be explained why the Qz excitations disappear in the AFM phase [20].

**Conclusion**

To summarize, we performed inelastic neutron scattering measurements on single crystals of $URu_{2-x}Re_xSi_2$ with x = 0.12, 0.15, and 0.25. Regardless of whether the electronic ground state is paramagnetic, gapless HO, or FM, no gap is observed in the magnetic excitations. This is consistent with the lack of specific heat anomalies in these concentrations and underscores that the entropy associated with the HO and AFM transitions in $URu_2Si_2$ is due to the gapping of the magnetic excitation spectrum, which itself is a feature of the hybridized band structure.

**Acknowledgments**

Access to MACS was provided by the Center for High Resolution Neutron Scattering, a partnership between the National Institute of Standards and Technology and the National Science Foundation under Agreement No. DMR-1508249. Sample synthesis at UCSD was supported by the U.S. Department of Energy, Office of Science, Office of Basic Energy Sciences under Award No. DE-FG02-04ER46105.

**References**

[1] M.B. Maple, J.W. Chen, Y. Dalichaouch, T. Kohara, C. Rossel, M.S. Torikachvili, and J.D. Thompson, Partially gapped Fermi surface in the heavy-electron superconductor uranium ruthenium silicide ($URu_2Si_2$), Phys. Rev. Lett. 56 (1986), pp. 185–188.

[2] T.T.M. Palstra, A.A. Menovsky, J.V.D. Berg, A.J. Dirkmaat, P.H. Kes, G.J. Nieuwenhuys, and J.A. Mydosh, Superconducting and magnetic transitions in the heavy-fermion system uranium ruthenium silicide ($URu_2Si_2$), Phys. Rev. Lett. 55 (1985), pp. 2727–2730.

[3] W. Schlabitz, J. Baumann, B. Pollit, U. Rauchschwalbe, H.M. Mayer, U. Ahlheim, and C.D. Bredl, Superconductivity and magnetic order in a strongly interacting Fermisystem: uranium ruthenium silicide ($URu_2Si_2$), Z. Phys. B Condens. Matter 62 (1986), pp. 171–177.

[4] J.A. Mydosh and P.M. Oppeneer, Colloquium: Hidden order, superconductivity, and magnetism: The unsolved case of $URu_2Si_2$, Rev. Mod. Phys. 83 (2011), pp. 1301–1322.

[5] K. Haule and G. Kotliar. Arrested Kondo effect and hidden order in $URu_2Si_2$, Nature Phys. 5 (2009) pp. 796-799.

[6] H. Ikeda, M.-T. Suzuki, R. Arita, T. Takimoto, T. Shibauchi and Y. Matsuda. Emergent rank-5 nematic order in $URu_2Si_2$, Nature Phys. 8 (2012) pp. 528-533.

[7] C. Broholm, H. Lin, P.T. Matthews, T.E. Mason, W.J.L. Buyers, M.F. Collins, A.A. Menovsky, J.A. Mydosh, and J.K. Kjems, Magnetic excitations in the heavy-fermion superconductor $URu_2Si_2$, Phys. Rev. B 43 (1991), pp. 12809–12822.

[8] C.R. Wiebe, J.A. Janik, G.J. MacDougall, G.M. Luke, J.D. Garrett, H.D. Zhou, Y.-J. Jo, L. Balicas, Y. Qiu, J.R.D. Copley, Z. Yamani, and W.J.L. Buyers, Gapped itinerant spin excitations account for missing entropy in the hidden-order state of $URu_2Si_2$, Nat. Phys. 3 (2007), pp. 96–99.

[9] F. Bourdarot, S. Raymond, and L.-P. Regnault. Neutron scattering studies on URu2Si2, Phil. Mag. 94 (2014), pp. 3702-3722.

[10] N.P. Butch, M.E. Manley, J.R. Jeffries, M. Janoschek, K. Huang, M.B. Maple, A.H. Said, B.M. Leu, and J.W. Lynn, Symmetry and correlations underlying hidden order in URu2Si2, Phys. Rev. B 91 (2015), p. 035128.

[11] Y. Dalichaouch, M. B. Maple, M. S. Torikachvili, and A. L. Giorgi, Ferromagnetic instability in the heavy-electron compound URu2Si2 doped with Re or Tc, Phys. Rev. B 39 (1989) pp. 2423-2431.

[12] M. Yokoyama, H. Amistuka, S. Itoh, I. Kawasaki, K. Tenya, and H. Yoshizawa, Neutron Scattering Study on Competition between Hidden Order and Antiferromagnetism in U(Ru1-xRhx)2Si2 (x ≤ 0.05), J. Phys. Soc. Jpn. 73 (2004) pp. 545-548.

[13] E.D. Bauer, V.S. Zapf, P.C. Ho, N.P. Butch, E.J. Freeman, C. Sirvent, and M.B. Maple, Non-Fermi-liquid behavior within the ferromagnetic phase in URu2−xRexSi2, Phys. Rev. Lett. 94 (2005), p. 046401.

[14] N.P. Butch and M.B. Maple, The suppression of hidden order and the onset of ferromagnetism in URu2Si2 via Re substitution, J. Phys. Condens. Matter 22 (2010), p. 164204.

[15] S. Ran, G. M. Schmiedeshoff, N. Pouse, I. Jeon, N. P. Butch, R. B. Adhikari, C. C. Almasan and M. B. Maple, Rapid suppression of the energy gap and the possibility of a gapless hidden order state in URu2−xRexSi2, Phil. Mag. 99 (2019) pp. 1751-1762.

[16] N.P. Butch and M.B. Maple, Evolution of critical scaling behavior near a ferromagnetic quantum phase transition, Phys. Rev. Lett. 103 (2009), p. 076404.

[17] J. A. Rodriguez, D. M. Adler, P. C. Brand, C. Broholm, J. C. Cook, C. Brocker, R. Hammond, Z. Huang, P. Hundertmark, J. W. Lynn, N. C. Maliszewskyj, J. Moyer, J. Orndorff, D. Pierce, T. D. Pike, G. Scharfstein, S. A. Smee and R. Vilaseca, MACS-a new high intensity cold neutron spectrometer at NIST, Meas. Sci. Technol. 19 (2008) p. 034023.

[18] J. R. Jeffries, N. P. Butch, B. T. Yukich, and M. B. Maple, Competing Ordered Phases in URu2Si2: Hydrostatic Pressure and Rhenium Substitution, Phys. Rev. Lett. 99 (2007) p. 217207.

[19] V. V. Krishnamurthy, D. T. Adroja, N. P. Butch, S. K. Sinha, and M. B. Maple, R. Osborn, J. L. Robertson, S. E. Nagler, and M. C. Aronson, Magnetic short-range

correlations and quantum critical scattering in the non-Fermi liquid regime of URu2−xRexSi2 (x=0.2–0.6), Phys. Rev. B **78**, (2008) p. 024413.

[20] N. P. Butch, S. Ran, I. Jeon, N. Kanchanavatee, K. Huang, A. Breindel, M. B. Maple, R. L. Stillwell, Y. Zhao, L. Harriger, and J. W. Lynn, Distinct magnetic spectra in the hidden order and antiferromagnetic phases in URu2−xFexSi2, Phys. Rev. B 94, (2016) p. 201102(R).

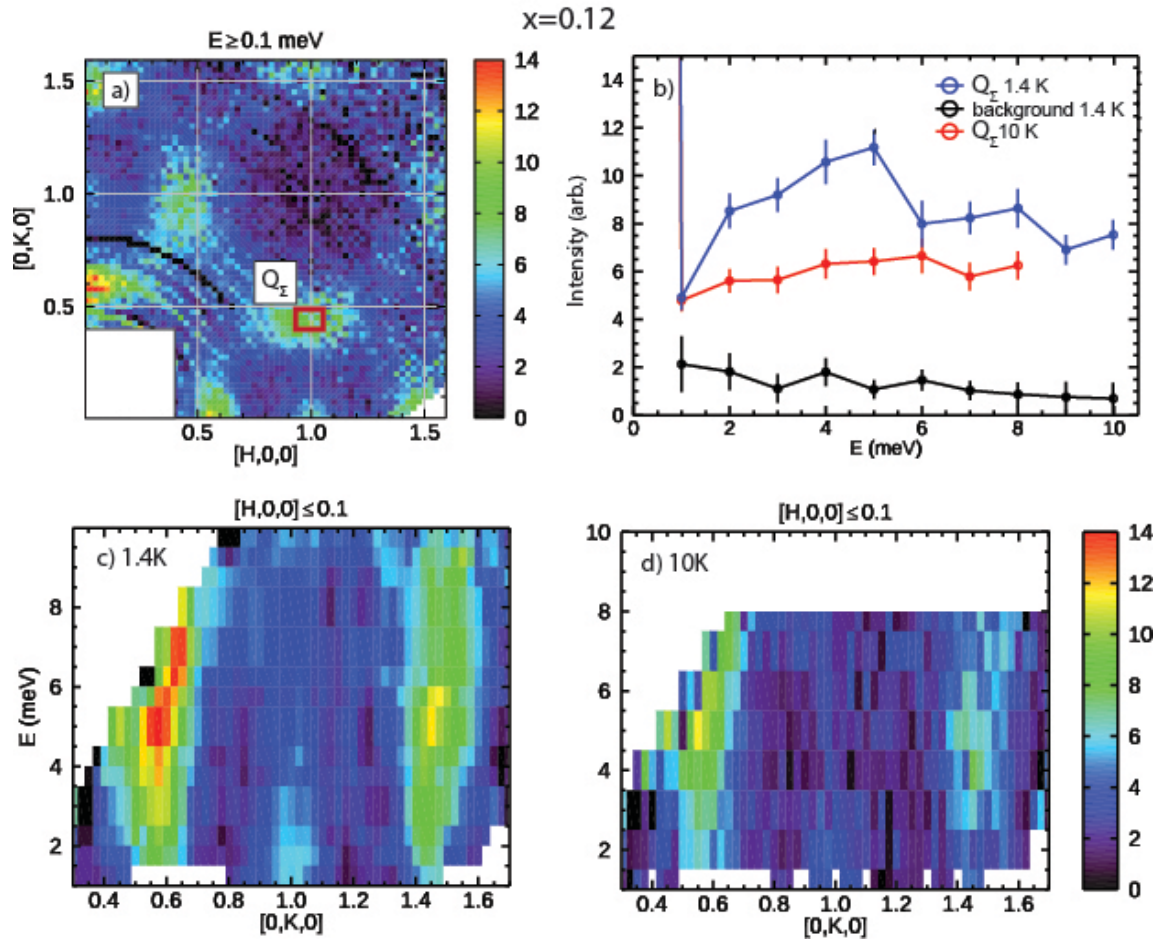

Figure 1. Magnetic excitations remain ungapped in the absence of HO in x=0.12. a) Basal-plane excitation distribution integrated over energy transfers up to 10 meV. b) The temperature- and E-dependence of the scattered intensity at $Q_\Sigma$, indicated in the red rectangle in a), showing that there is no energy gap. The signal is well above background. c,d) The Q-dependence of the magnetic excitations at 1.4K and 10K, showing how all excitations remain ungapped.

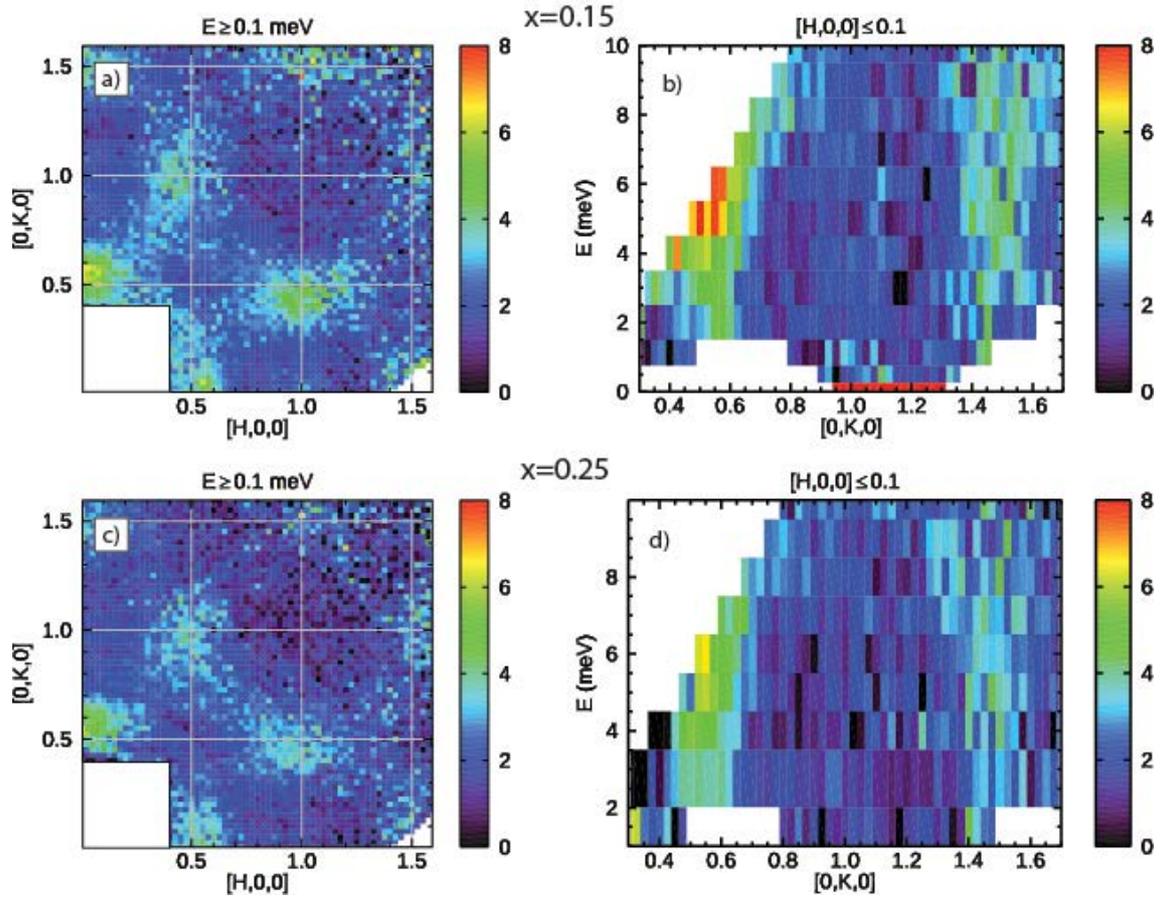

Figure 2. Ungapped magnetic excitations beyond the HO phase in x=0.15 and x=0.25 at 1.4 K. a) The Q-dependence at the nominal quantum critical point (x=0.15) remains the same, while the b) incommensurate excitations are ungapped. c,d) The Q- and E-dependence are also similar in the FM state (x=0.25).